\documentclass{nature}
\usepackage{xcolor}
\usepackage{amsmath,amssymb}
\usepackage{graphicx}
\usepackage[utf8]{inputenc}
\usepackage[normalem]{ulem} 

\newcounter{lastnote}
\newenvironment{scilastnote}{%
\setcounter{lastnote}{\value{enumiv}}%
\addtocounter{lastnote}{+1}%
\begin{list}%
{\arabic{lastnote}.}
{\setlength{\leftmargin}{.22in}}
{\setlength{\labelsep}{.5em}}}
{\end{list}}

\title{Ultrastrong Magnon-Magnon Coupling Dominated by \\Antiresonant Interactions}

\author
{Takuma Makihara,$^{1}$ Kenji Hayashida,$^{2,3}$ G. Timothy Noe II,$^{2}$ Xinwei Li,$^{2}$ Nicolas Marquez Peraca,$^{1}$ \\Xiaoxuan Ma,$^{4}$ Zuanming Jin,$^{5}$ Wei Ren,$^{4}$ Guohong Ma,$^{4}$ Ikufumi Katayama,$^{6}$ Jun Takeda,$^{6}$ \\Hiroyuki Nojiri,$^{7}$ Dmitry Turchinovich,$^{8}$ Shixun Cao,$^{4}$ Motoaki Bamba,$^{9,10}$ Junichiro Kono$^{1,2,11}$} 

\begin{document}

\maketitle 


\begin{affiliations}
	\item Department of Physics and Astronomy, Rice University, Houston, Texas, USA
	\item Department of Electrical and Computer Engineering, Rice University, Houston, Texas, USA
	\item Division of Applied Physics, Graduate School and Faculty of Engineering, Hokkaido University, Sapporo, Hokkaido 060-8628, Japan
	\item Department of Physics, International Center of Quantum and Molecular Structures and Materials Genome Institute, Shanghai University 200444, Shanghai, China
	\item Terahertz Technology Innovation Research Institute, Terahertz Spectrum and Imaging Technology Cooperative Innovation Center, Shanghai Key Lab of Modern Optical System, University of Shanghai for Science and Technology, 516 JunGong Road, Shanghai 200093, China
	\item Department of Physics, Graduate School of Engineering Science, Yokohama National University, Yokohama, Japan
	\item Institute for Materials Research, Tohoku University, Sendai, Japan
	\item Fakult\"{a}t f\"{u}r Physik, Universit\"{a}t Bielefeld, Universit\"{a}tsstr. 25, 33615 Bielefeld, Germany
	\item Department of Physics, Kyoto University, Sakyo-ku, Kyoto, 606-8502, Japan
	\item PRESTO, Japan Science and Technology Agency, Saitama 332-0012, Japan
	\item Department of Materials Science and NanoEngineering, Rice University, Houston, Texas, USA
\end{affiliations}




\date{}






\newpage


\begin{abstract}
    Exotic quantum vacuum phenomena are predicted in cavity quantum electrodynamics (QED) systems with ultrastrong light-matter interactions. Their ground states are predicted to be vacuum squeezed states with suppressed quantum fluctuations. The source of such phenomena is antiresonant terms in the Hamiltonian, yet antiresonant interactions are typically negligible compared to resonant interactions in light-matter systems. We report an unusual, ultrastrongly coupled matter-matter system of magnons that is analytically described by a unique Hamiltonian in which the relative importance of resonant and antiresonant interactions can be easily tuned and the latter can be made vastly dominant. We found a novel regime where vacuum Bloch-Siegert shifts, the hallmark of antiresonant interactions, greatly exceed analogous frequency shifts from resonant interactions. Further, we theoretically explored the system’s ground state and calculated up to 5.9 dB of quantum fluctuation suppression. These observations demonstrate that magnonic systems provide an ideal platform for exploring exotic quantum vacuum phenomena predicted in ultrastrongly coupled light-matter systems.
\end{abstract}

\newpage

The interaction of light with solids can exhibit extreme values of coupling strength, unachievable in atomic and molecular systems, due to large dipole moments and cooperative many-body interactions characteristic of condensed matter.
For example, Dicke cooperativity,\cite{Dicke54PR} a quantum optical phenomenon where $N$ dipoles coupled to a single electromagnetic field experience a light-matter coupling strength enhanced by a factor of $\sqrt{N}$, becomes drastic in condensed matter. 
Leveraging cooperative many-body interactions enables observations of the exotic ultrastrong coupling (USC) regime.\cite{FornDiazetAl19RMP,KockumetAl19NRP}

In the USC regime, the light-matter coupling strength becomes comparable to the bare frequencies of the system. In this regime, the rotating-wave approximation (RWA) breaks down, leading to antiresonant interactions from the so-called counter-rotating terms (CRTs) and $A^2$ terms in the Hamiltonian, which allow access to counter-intuitive and unexplored physics. In the past decade, USC has been realized in diverse physical platforms, including intersubband polaritons,\cite{TodorovetAl10PRL,GeiseretAl12PRL} Landau polaritons,\cite{ScalarietAl12Science,ZhangetAl16NP,BayeretAl17NL} and superconducting circuits.\cite{NiemczyketAl10NP,Forn-DiazetAl10PRL,YoshiharaetAl17NP,Forn-DiazetAl17NP} 
However, traditional polariton systems are restricted by one fixed coupling strength, and resonant effects, such as vacuum Rabi splitting (VRS), dominate antiresonant effects, such as vacuum Bloch-Siegert shifts (VBSSs),\cite{BlochSiegert40PR} which are the hallmark of active CRTs. Thus, the counter-intuitive physics predicted in this regime, such as the superradiant phase transition,\cite{HeppLieb73AP,WangHioe73PRA} Casimir photon emission,\cite{DeLiberatoetAl07PRL,StassietAl13PRL,Hagenmuller16PRB} and ground-state electrolumin-\\escence,\cite{CirioetAl16PRL} has largely remained unexplored. 
Experimental studies are largely limited to reports of giant VRS, and there have only been a few unambiguous demonstrations of the VBSS.\cite{LietAl18NP} 
Therefore, there is a growing demand for novel platforms with superior tunability and dominant antiresonant interactions for exploring the exotic predictions of the USC regime.

Here, we demonstrate novel matter-matter USC in YFeO$_3$, a rare-earth orthoferrite,\cite{White69JAP} that is analytically described by a unique cavity QED Hamiltonian with tunable coupling strengths and dominant counter-rotating interactions. We systematically examined how the quasi-ferromagnetic (qFM) and quasi-antiferromagnetic (qAFM) magnon modes modes interact with each other by characterizing their resonance frequencies at different applied magnetic field strengths and directions. We were able to use the applied magnetic field to tune the VRS and VBSSs, and in certain geometries, the frequency shifts of the coupled modes were dominated by the VBSSs and not the vacuum Rabi splitting-induced shifts (VRSSs). A well-established microscopic spin model of this material system\cite{Herrmann63JPCS} accurately reproduced our observed resonances without any adjustable parameters. We show that this lightless spin model can be precisely mapped to a polariton model described by an \emph{anisotropic} Hopfield Hamiltonian in which the magnon-magnon coupling strengths are easily tunable and the CRTs dominate the co-rotating terms, consistent with our observation of giant VBSSs. Finally, we theoretically investigated the ground state of our system and demonstrate that it is intrinsically squeezed, consisting of a two-mode squeezed vacuum as expected in the USC regime,\cite{Artoni-Birman89QO,Schwendimann-Quattropani92EPL,CiutietAl05PRB} with quantum fluctuation suppression as large as $5.9$~dB. 

\section*{Results}
\textbf{Terahertz time-domain magnetospectroscopy}. To interrogate magnons in YFeO$_3$, we used terahertz time-domain spectroscopy (THz TDS). In THz TDS studies of rare-earth orthoferrites, free-induction decay signals from precessing spins are measured directly in the time domain, the Fourier transform of which reveal the precessional (magnon) frequency.\cite{YamaguchietAl10PRL} 
%
We combined two unique experimental apparatuses: a table-top, 30~T pulsed magnet\cite{NoeetAl13RSI} and single-shot THz detection,\cite{NoeetAl16OE,MinamietAl13APL} illustrated in Figure~1a. 
THz pulses were focused onto the samples, and the transmitted THz waveform was detected using 
a single-shot technique based on a reflective echelon that separates an optical probe pulse into time-delayed beamlets that overlap with the THz waveform in our ZnTe detection crystal.\cite{NoeetAl16OE,MinamietAl13APL}  Figure~1b displays a THz waveform transmitted through YFeO$_3$ and detected using single-shot detection. Coherent oscillations are observed for $t>0$, whose Fourier transform reveals the magnon frequency (inset). Figure~1c shows the magnetic field profile, the detection optical pulses, and the sampled magnetic field strengths, as well as the THz waveforms measured at the sampled field strengths.

\noindent \textbf{Demonstration of ultrastrong magnon-magnon coupling}. In the absence of an applied external magnetic field ($\mathbf{H}_{\mathrm{DC}}=0$), YFeO$_3$ crystallizes in an orthorhombic perovskite structure (see Figure~3b). Its magnetic structure is described by the $\Gamma_4$ phase, where the two Fe$^{3+}$ spin sublattices ($\mathbf{S}_1$ and $\mathbf{S}_2$) order antiferromagnetically along the $a$-axis, with a slight canting towards the $c$-axis due to the Dzyaloshinskii-Moriya interaction. Figure~2 shows results of THz magnetospectroscopy studies of YFeO$_3$. We studied five different single crystals of YFeO$_3$ cut such that the applied magnetic field, $\mathbf{H}_{\mathrm{DC}}$, was directed at angles of $\theta = 0^{\circ}, 20^{\circ}, 40^{\circ}, 60^{\circ}$, and $90^{\circ}$ with respect to the \textit{c}-axis in the \textit{b-c} plane. The measurements were conducted at room temperature in the geometry shown in Figure~2a. The THz radiation propagated parallel to $\mathbf{H}_{\mathrm{DC}}$, and the incident THz electric field $\mathbf{E}_{\mathrm{THz}}$ was linearly polarized along the \textit{a}-axis. In general, the emitted THz electric fields were elliptically polarized,\cite{YamaguchietAl13PRL} so THz electric fields polarized parallel to the $a$-axis ($\mathbf{E}_{\mathrm{THz}}^{a}$) and polarized in the $b$-$c$ plane ($\mathbf{E}_{\mathrm{THz}}^{b-c}$) were both measured to fully characterize the magnetic resonances.\cite{YamaguchietAl10PRL,YamaguchietAl13PRL} 

Figure~2b displays an example THz waveform and its Fourier transform for an applied field strength of $12.60$~T at $\theta = 20^{\circ}$. Beating in the time domain, and correspondingly two peaks in the frequency domain, indicate the simultaneous excitation of two magnon modes. Figure~2c displays an example of the two magnon frequencies extracted by Fourier transforming $\mathbf{E}_{\mathrm{THz}}^{b\textrm{-}c}$ for $\theta = 20^{\circ}$ and $40^{\circ}$ at different magnetic fields (see Methods for all measurements). Figure~3a plots the observed resonance frequencies (black dots) versus magnetic field for all measured $\theta$. We observe anticrossing between the two frequencies whose splitting increases with increasing $\theta$, illustrating strong coupling between the two magnons with tunable coupling strengths. Further, the frequency splitting is comparable to the bare magnon frequencies, indicating ultrastrong magnon-magnon coupling.

We first interpret our observed magnon-magnon coupling by considering the symmetry of the spin dynamics in the qFM and qAFM modes. Figure~3b displays the crystal and magnetic structure of YFeO$_3$. Figure~3c qualitatively illustrates the spin precessions in the qFM and qAFM modes for $\mathbf{H}_{\mathrm{DC}} = 0$. In this geometry, or when $\mathbf{H}_{\mathrm{DC}}$ is applied along the $c$-axis ($\theta = 0^\circ$), $\mathbf{S}_1$ and $\mathbf{S}_2$ maintain $\pi$ rotational symmetry about the $c$-axis, and the qFM and qAFM modes do not hybridize due to opposite parities under this symmetry: the qAFM mode is unchanged whereas the qFM mode gains a $\pi$ phase shift.\cite{WhiteetAl82PRB,MacNeilletAl19PRL} However, this symmetry is broken when the spins possess a component along the $b$-axis, allowing the qFM and qAFM modes to hybridize. We employ a tilted $\mathbf{H}_{\mathrm{DC}}$ in the $b$-$c$ plane to prepare an equilibrium spin configuration where $\mathbf{S}_1$ and $\mathbf{S}_2$ posses components along the $b$-axis, as in Figure~3b, and enable hybridization.

\noindent \textbf{Microscopic Spin Model}. To quantitatively illustrate the hybridization between the qFM and qAFM modes, we numerically analyzed the spin dynamics in a tilted magnetic field. We started from a microscopic spin model describing interactions between $\mathbf{S}_1$ and $\mathbf{S}_2$, including the symmetric exchange, the antisymmetric exchange, the single-ion anisotropies, and the Zeeman interaction.\cite{Herrmann63JPCS} The model contains no fitting parameters; the inputted magnetic parameters are well-known for YFeO$_3$.\cite{KoshizukaHayashi88JPSJ} By solving the Landau-Lifshitz-Gilbert equation, we obtained the spin dynamics in a tilted magnetic field. Figure~3a plots the resonance frequencies (solid red lines), which excellently reproduce our experimental results (black dots). We observe that for nonzero $\theta$, the resonance frequencies display anticrossing, indicating mode hybridization consistent with the broken $\pi$ rotational symmetry mentioned above. The two coupled modes are labeled as the upper mode (UM) and lower mode (LM) for the higher and lower frequency branches, respectively. We do not excite the qAFM mode when $\theta = 0^\circ$ and we do not excite the LM when $\theta = 90^\circ$ due to magnon excitation selection rules (see Methods). 

We calculated the dynamics of the decoupled qFM and qAFM modes in a tilted magnetic field, which are uniquely defined by opposite parities under $\pi$ rotation about the $c$-axis, by neglecting coupling between these independent spin precessions in the equations of motion. The decoupled magnon frequencies are plotted as black dashed-dotted lines in Figure~3a for nonzero $\theta$. Note that for $\theta = 0^\circ$, the qFM and qAFM modes solve the full equations of motion.
For $\theta = 20^\circ$, $40^\circ$, and $60^\circ$, we observe that the UM is higher in frequency than the qFM and qAFM modes. This is precisely what one expects for hybridization within the RWA; for the UM, the VRSS (exclusively from the co-rotating interaction) is always a blue-shift.
However, we observe that for $\theta = 90^\circ$, the UM is lower in frequency than the qAFM mode, indicating an additional red-shift of the coupled magnon frequencies. This dominant red-shift, which is a direct consequence of the counter-rotating term, is the dominant VBSS. Due to giant VBSSs, we observe that the UM and LM, whose dynamics are qualitatively illustrated in Figure~3c for $\theta = 90^\circ$, not only hybridize the qFM and qAFM modes but also contain the time-reversed dynamics of the qFM and qAFM modes (see Methods). Additional calculations showing the transition from $\theta = 60^\circ$ to $\theta = 90^\circ$ are shown in Supplementary Fig. 11.

\noindent \textbf{Quantum Mechanical Model}. To evaluate the magnon-magnon coupling strengths, we rewrite our microscopic spin model in terms of the creation and annihilation operators of the qFM and qAFM magnons:
\begin{equation}
    \mathcal{H} = \hbar\omega_{0a} \left (\hat{a}^\dagger\hat{a}+\frac{1}{2} \right ) + \hbar\omega_{0b} \left ( \hat{b}^\dagger\hat{b} + \frac{1}{2} \right ) + i\hbar g_1 \left ( \hat{a}\hat{b}^\dagger - \hat{a}^\dagger\hat{b} \right ) + i\hbar g_2 \left ( \hat{a}^\dagger \hat{b}^\dagger - \hat{a}\hat{b} \right )
    \label{ref: Ham}
\end{equation}
where $\hat{a}$ ($\hat{a}^\dagger$) annihilates (creates) a qFM magnon with frequency $\omega_{0a}$, and $\hat{b}$ ($\hat{b}^\dagger$) annihilates (creates) a qAFM magnon with frequency $\omega_{0b}$, where $\omega_{0a}$ and $\omega_{0b}$ are the frequencies of the decoupled qFM and qAFM modes discussed in the previous paragraph. Expressions for the co-rotating coupling strength ($g_1$) and the counter-rotating coupling strength ($g_2$), which are derived in the absence of adjustable parameters, are provided in the Methods. 
Our Hamiltonian resembles the Hopfield Hamiltonian,\cite{Hopfield58PR} which is related to the paradigmatic Dicke Hamiltonian by a Holstein-Primakoff transformation. However, unlike the Hopfield Hamiltonian and analogous light-matter Hamiltonians, such as the quantum Rabi model or the Dicke Hamiltonian, our system is not restricted to $g_1 = g_2$. Although similar anisotropic Hamiltonians, such as the anisotropic quantum Rabi model, have been theoretically proposed\cite{XieetAl14PRX} and experimentally realized in superconducting circuits,\cite{LuetAl17PRL2} our condensed matter system can simulate many-body Hamiltonians. For example, the Hopfield Hamiltonian is typically used in studies of USC in condensed matter systems, such as intersubband polaritons\cite{CiutietAl05PRB} and exciton-polaritons.\cite{Kena-CohenetAl13AOM,GambinoetAl14ACS}

Figure~4a plots values of $|g_1|$ and $|g_2|$ versus applied magnetic field for $\theta = 20^{\circ}, 40^{\circ}, 60^{\circ}$, and $90^{\circ}$ (see Supplementary Fig. 13 for plots of $|g_1|$ and $|g_2|$ versus $\theta$ for different applied magnetic fields). The case where $\theta = 0^{\circ}$ is not shown because $g_1$ and $g_2$ exactly vanish in this geometry. 
Importantly, Figure~4a demonstrates tunable, anisotropic co-rotating and counter-rotating coupling strengths, with the latter always dominating the former, indicating an extreme breakdown of the RWA. This observation builds on recent experiments\cite{liensberger2019exchange} to demonstrate that magnonic systems can not only achieve tunable ultrastrong coupling, but antiresonant ultrastrong coupling with tunable anisotropy, in a material compatible with ultrafast coherent control. Further, we observed that $|g_2|$ monotonically increase with $\theta$. We found that the anisotropy between $g_1$ and $g_2$ depends on magnetic parameters through the spin canting angle, and that the coupling becomes more anisotropic as the canting angle decreases (see Supplementary Text, Supplementary Fig. 12).
For $\theta = 60^\circ$ and $90^\circ$, Figure~4b plots the qFM and qAFM modes (black dashed-dotted lines), the LM and UM (red solid lines), and the co-rotating coupled magnon frequencies (green dashed lines) that are obtained by setting $g_2 = 0$. The VBSSs are indicated by the shaded areas between the red solid lines and the green dashed lines. We see that for $\theta = 60^\circ$, the VBSSs are small relative to the VRSSs (differences between green dashed lines and black dashed-dotted lines), but that the opposite is true when $\theta = 90^\circ$. For the UM, the VBSS even becomes dominant when $\theta = 90^\circ$, consistent with the increase of $|g_2|$ relative to $|g_1|$ with increasing $\theta$. Our observation of a dominant VBSS for the UM is unique to the anisotropic Hopfield Hamiltonian and can only be achieved for $|g_2| > |g_1|$ (see Methods).
Figure~4c plots figures of merit referred to as normalized coupling strengths for $0^\circ \leq \theta \leq 89^\circ$ (the case for $90^\circ$ is discussed in Methods). The normalized coupling strength, defined as the ratio of the coupling strength to the frequency where the decoupled qFM and qAFM modes cross ($\omega_0$), determines whether a system is in the USC regime.\cite{FornDiazetAl19RMP,KockumetAl19NRP} In a system characterized by one coupling strength $g$, the USC regime has been defined as when $g/\omega_0 > 0.1$. Thus, we observe that our system can be continuously tuned between no coupling and USC as a function of $\theta$, with the maximum experimentally accessible normalized coupling strengths occurring at $\theta = 58^\circ$ for 30~T, and are given by $|g_1|/\omega_0 = 0.26$ and $|g_2|/\omega_0 = 0.39$. 

Our observation of large counter-rotating interactions is expected to amplify the two-mode vacuum squeezing of the ground state that was discussed in the earliest theoretical study of USC in an isotropic Hopfield Hamiltonian.\cite{CiutietAl05PRB} To demonstrate the capabilities of using YFeO$_3$ to realize a magnonic two-mode squeezed vacuum, we evaluate the quantum fluctuations in our system. We first define a generalized magnon annihilation operator $\hat{c} = \alpha\hat{a} + \beta\hat{b}$ and its corresponding quadrature $\hat{X}_{\hat{c},\phi} = (\hat{c}e^{i\phi}+\hat{c}^{\dagger}e^{-i\phi})/2$. The standard quantum limit for the fluctuation of $\hat{X}_{\hat{c},\phi}$, defined as its variance when evaluated in the decoupled magnon vacuum, is given by $1/4$. We numerically investigated the minimum fluctuation in $\hat{X}_{\hat{c},\phi}$ evaluated in the ground state of our coupled magnon system and observed a clear suppression below the standard quantum limit.

Figure~4d illustrates this fluctuation suppression below the standard quantum limit of $0$~dB, where we numerically searched for the parameters $\alpha$, $\beta$, and $\phi$ that minimize this fluctuation. An orthogonal operator to $\hat{c}$ also demonstrating squeezing is discussed in the Methods section. We observed that the maximum experimentally achievable squeezing is $5.9$~dB, which occurs at $30$~T for $\theta = 90^\circ$. This strong degree of squeezing is a direct consequence of our large CRTs. 
Figure~4d demonstrates that the degree of squeezing in our system is easily tunable with applied magnetic field strength and direction, going beyond previous works studying antiferromagnetic magnon squeezing due soley to intrinsic material properties.\cite{kamra2019antiferromagnetic} We note that the squeezing is related to magnetic parameters through $g_2$ as the counter-rotating interaction is the source of squeezing. Therefore, increasing the anisotropy between $g_2$ and $g_1$ (see Supplementary Text, Supplementary Fig. 12), leading to stronger counter-rotating interactions, could further amplify the squeezing. We also observed that the quadrature fluctuations approach zero as $\theta$ approaches $90^\circ$ when evaluated at the field strength where the qFM and qAFM modes cross (see Supplementary Fig. 10), suggesting that our system reaches a critical point. We numerically found that complete quadrature fluctuation suppression (i.e. perfect squeezing) is obtained at a critical coupling strength above which the LM becomes gapless, suggesting a magnonic superradiant phase transition. 

\section*{Conclusions}
\indent Our observation of tunable, anisotropic coupling strengths with the CRTs dominating the co-rotating terms demonstrates that magnons in rare-earth orthoferrites serve as an ideal platform for studying many-bodied quantum optical phenomena in extreme regimes of coupling strengths that are inaccessible to traditional photonic systems. In particular, the magnonic ground state describable as a two-mode squeezed vacuum may lead to a pathway for decoherence-free quantum information technology. Perfect magnon squeezing, predicted for a magnonic superradiant phase, will produce a novel platform of many-body physics to explore the correlation between the quantum phase transitions and the exotic quantum fluctuations.


\newpage


\newpage

\begin{scilastnote}
\newpage


\item[{\bf \,\,\,Figures}]
\hfill \\
\begin{figure}
    \centering
    \includegraphics[width = 3.5 in]{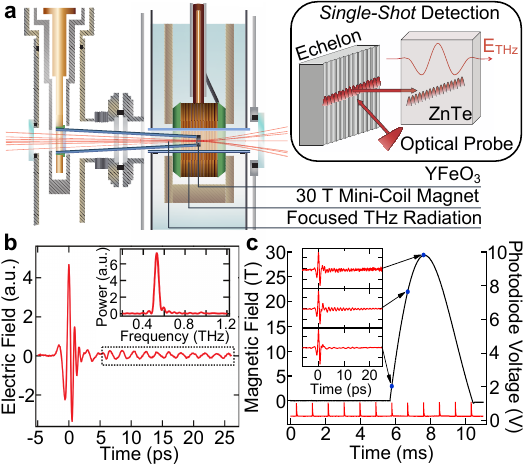}
\end{figure}

{\bf Fig.~1.} \textbf{Unique combination of table-top, 30 T pulsed magnet and single-shot THz detection.} \textbf{a},~Schematic of pulsed magnet surrounding single crystals of YFeO$_3$. Note that although the sample is held on a sapphire pipe mounted on the cold finger of a liquid helium cryostat, no liquid helium was used in this study. Single-shot detection (shown in inset) is based on a unique reflective echelon. \textbf{b},~Sample THz electric field waveform transmitted through a YFeO$_3$ crystal. Time-domain oscillations for $t>0$ from coherent spin precessions (magnons) are Fourier transformed to yield the magnon frequency (inset). \textbf{c},~Pulsed magnetic field profile (solid black line), optical pulses used to generate/detect THz waveforms (solid red line), and sampled magnetic field strengths (blue dots) with transmitted THz waveforms measured at the sampled magnetic field strength shown in the inset. The optical pulses are detected using a photodiode that measures scattered light.

\newpage
\begin{figure}
    \centering
    \includegraphics[width = 3.5 in]{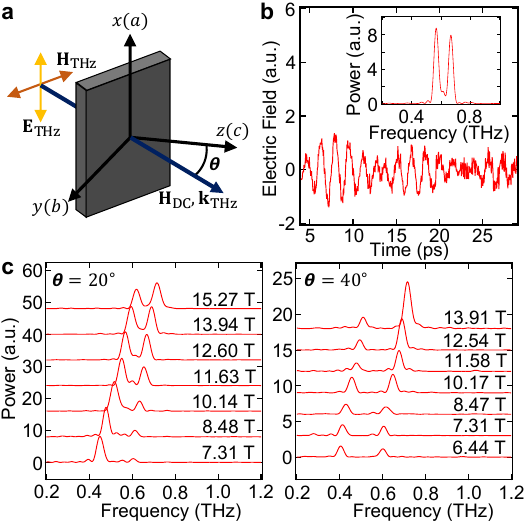}
\end{figure}

{\bf Fig.~2.} \textbf{Magnon signals in time and frequency domains.} \textbf{a},~Schematic of THz magnetospectroscopy studies of YFeO$_3$ in a tilted magnetic field. $\mathbf{H}_\text{DC}$ was applied in the $b-c$ plane at an angle of $\theta$ with respect to the $c$-axis, with $\mathbf{k}_\text{THz}//\mathbf{H}_\text{DC}$ and $\mathbf{H}_\text{THz}$ polarized in the $b-c$ plane. \textbf{b},~Transmitted THz waveform for $\theta = 20^{\circ}$ at $\mathbf{H}_\text{DC} = 12.60$ T displaying beating in the time-domain and two peaks in the frequency domain corresponding to the simultaneous excitation of both magnon modes in YFeO$_3$. \textbf{c},~Magnon power spectra for $\theta = 20^{\circ}$ and $\theta = 40^{\circ}$ at different $\mathbf{H}_\text{DC}$ displaying larger frequency splitting for larger $\theta$.

\bigskip
\begin{figure}
    \centering
    \includegraphics[width = 1\linewidth]{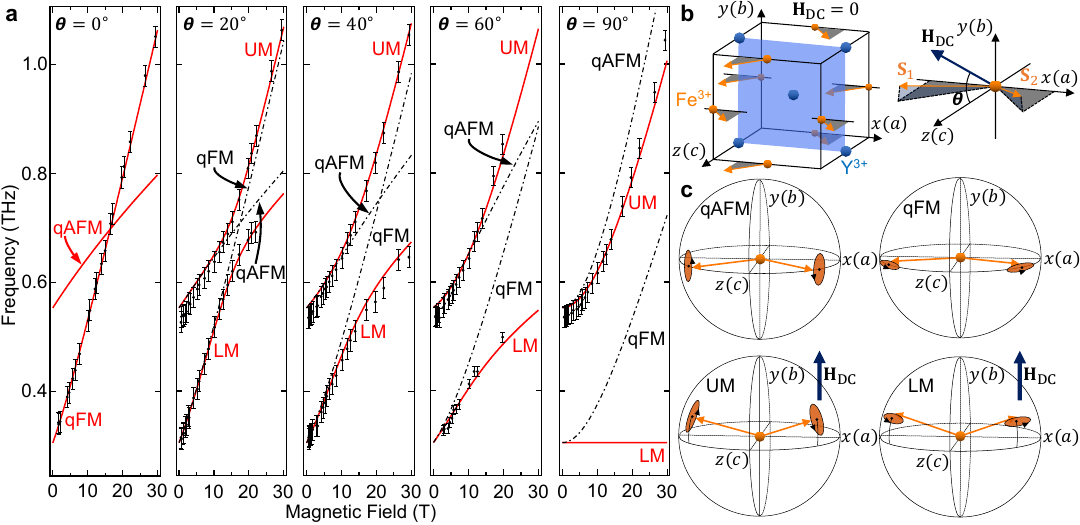}
\end{figure}

{\bf Fig.~3.} \textbf{Evidence for dominant vacuum Bloch-Siegert shifts.} \textbf{a},~Experimentally measured magnon frequencies for $\theta = 0^{\circ}, 20^{\circ}, 40^{\circ}, 60^{\circ}, 90^{\circ}$ versus $\mathbf{H}_{\mathrm{DC}}$ (black dots) with calculated resonance magnon frequencies (solid red lines) and decoupled qFM and qAFM magnon frequencies (black dashed-dotted lines). The UM frequency becomes lower than the qAFM frequency at $\theta = 90^{\circ}$, indicating a dominant VBSS compared to the VRSS. Error bars are $1/T$ where $T$ is the Fourier-transformed time-window and is limited by our THz detection. \textbf{b},~Crystal and magnetic structure of YFeO$_3$. \textbf{c},~Spin dynamics in the decoupled qFM and qAFM modes, as well as the UM and LM.

\newpage
\begin{figure}
     \centering
     \includegraphics[width = 0.95\linewidth]{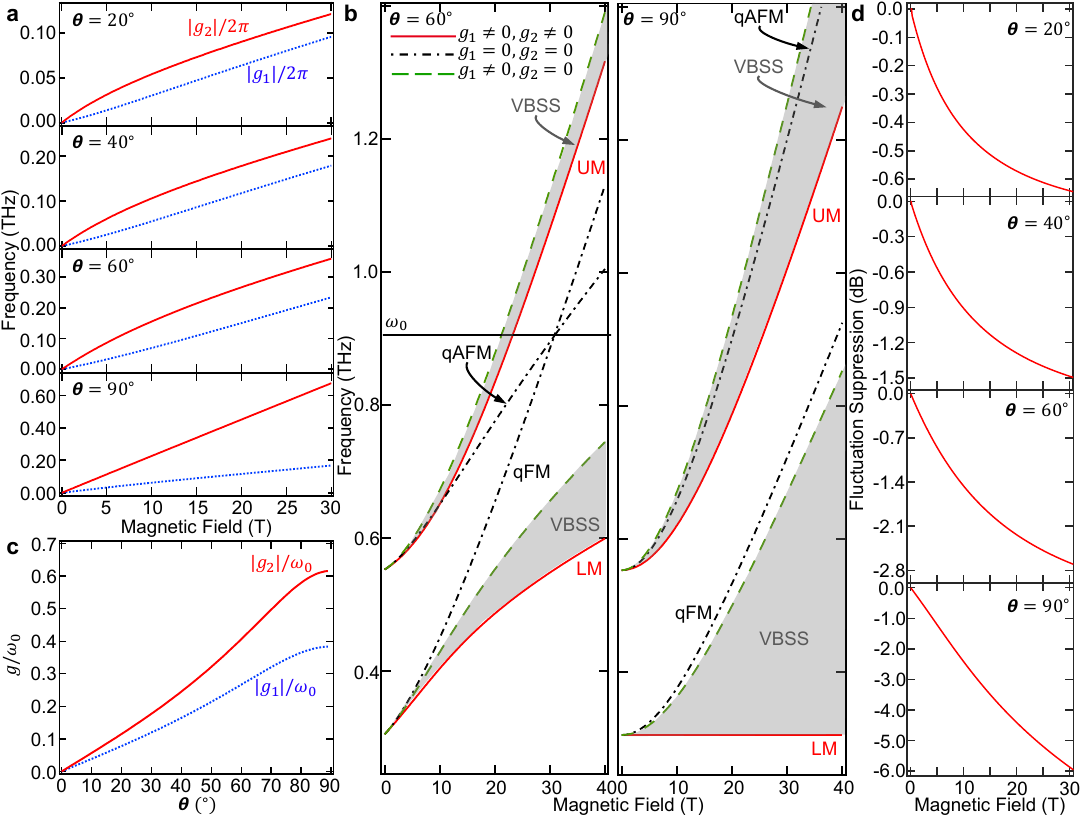}
\end{figure}

{\bf Fig.~4.} \textbf{Extreme breakdown of rotating-wave approximation and vacuum squeezing.} \textbf{a},~Co-rotating ($|g_1|/2\pi$, blue dotted line) and counter-rotating ($|g_2|/2\pi$, red solid line) coupling strengths for $\theta = 20^{\circ}, 40^{\circ}, 60^{\circ}, 90^{\circ}$ displaying dominance of the counter-rotating term. \textbf{b},~Theoretical illustration of qFM mode, qAFM mode, LM, UM, and co-rotating coupled magnon frequencies that are obtained by setting $g_2 = 0$, for $\theta = 60^{\circ}$ and $90^{\circ}$. The VBSSs are highlighted by the shaded area. \textbf{c},~Normalized co-rotating ($|g_1|/\omega_0$, blue dotted line) and counter-rotating ($|g_2|/\omega_0$, red solid line) coupling strengths displaying ultrastrong magnon-magnon coupling and dominance of the CRTs. $\omega_0$ is the frequency at which the qFM and qAFM modes cross, as illustrated in \textbf{b}. \textbf{d},~Fluctuation suppression in $\hat{X}_{\hat{c},\phi}$ evaluated in the ground state of the coupled magnon system for $\theta = 20^{\circ}, 40^{\circ}, 60^{\circ}$, and $90^{\circ}$ demonstrating squeezing.  For $\theta = 90^{\circ}$, suppression reaches $5.9$~dB for $30$~T.


\newpage


\item[{\bf \,\,\,Methods}]
\hfill \\
{\bf Sample Preparation}\\
Polycrystalline YFeO$_3$ was synthesized by conventional solid state reaction using Y$_2$O$_3$ ($99.9$\%) and Fe$_2$O$_3$ ($99.9$\%) powders. According to stoichiometric ratios, original reagents were weighted and pulverized with moderate anhydrous ethanol in an agate mortar. Mixtures were sintered at $1,300$ $^\circ$C for $1,000$ minutes, then furnace cooled down to room temperature. We continued to grind the presintered sample into powder, pressed it into sheets, reduced the gap between the powder particles, and conducted the second sintering. The sintering temperature and duration were the same as the pre-firing process. The secondary sintered pellets were thoroughly reground, and the polycrystalline powders were pressed into a rod that is $70$-$80$~mm in length and $5$-$6$~mm in diameter by a Hydrostatic Press System under $70$~MPa, and then sintered again at $1,300$ $^\circ$C.

Single crystals were grown in the optical floating zone furnace (FZT-10000-H-VI-P-SH, Crystal Systems Corp; Heat source: four $1$~kW halogen lamps). During the crystal growth process, we used a YFeO$_3$ single crystal as a seed crystal. The molten zone moved upwards at a rate of $3$~mm/h with the seed rod (lower shaft) and the feed rod (upper shaft) counter rotating at $30$~rpm in airflow by $3$~L/min.

We characterized our obtained crystals with a back-reflection Laue camera and X-ray diffraction (XRD). The results show that the sample is a high quality single crystal without impurity phase. We further prepared sheet samples of YFeO$_3$ single crystals along the three crystal axis directions for XRD measurement to ensure the accuracy of the crystal directions. 

{\bf THz Time-Domain Magnetospectroscopy}\\
We performed THz time-domain magnetospectroscopy by combining $30$-T pulsed magnetic fields with THz TDS. The output from an amplified Ti:Sapphire laser ($1$~kHz, $150$~fs, $775$~nm, $0.8$~mJ, Clark-MXR, Inc., CPA-2001) is divided between THz generation and detection paths. Intense THz is generated using the tilted-pulse-front excitation method\cite{HeblingetAl08JOSAB} in LiNbO$_3$ and is detected using free-space electro-optic (EO) sampling in ZnTe. The incident THz electric field was linearly polarized parallel to the $a$-axis using a wire-grid polarizer. Transmitted THz electric field components parallel ($\mathbf{E}_{\mathrm{THz}}^{a}$) and perpendicular ($\mathbf{E}_{\mathrm{THz}}^{b\textrm{-}c}$) to the incident radiation were identified using a second wire-grid polarizer, then focused onto the detection crystal.

Magnetic fields up to $30$~T were generated in the Rice Advanced Magnet with Broadband Optics (RAMBO), a table-top pulsed magnet that combines strong magnetic fields with diverse spectroscopies.\cite{NoeetAl13RSI} A schematic of RAMBO is illustrated in Supplementary Fig.\,1 and is discussed in the Supplementary Text. 

Because our magnetic field changes with time, we must rapidly sample the entire THz waveform. We achieve this by implementing single-shot THz detection using a reflective echelon that separates a reflected optical probe pulse into time-delayed beamlets, thereby stretching the optical pulse front.\cite{NoeetAl16OE} This linearly polarized stretched pulse front overlaps with the entire THz waveform in our detection crystal. The detection crystal is followed by a quarter-wave plate, a Wollaston prism, and imaging optics to separate orthogonal polarization components of the probe, which we use to generate two images of the reflective echelon on a CMOS camera. Supplementary Fig.\,2a displays images of the reflective echelon without (top) and with (bottom) THz radiation propagating through the detection crystal. The red dashed box highlights the position of the large-amplitude THz electric field pulse corresponding to $t=0$ in the time-domain. We describe how we obtain the THz electric field from these images in the Supplementary Text. 

For quantitative measurements of the THz electric field, we analyze images both in the presence and absence of the THz electric field, yielding a \textit{signal} and a \textit{reference}, respectively. Supplementary Fig.\,2b displays the signal and reference obtained from the echelon images in Supplementary Fig.\,2a, as well as the \textit{waveform}, obtained by taking the difference of the signal and the reference. 

{\bf Magnon Mode Excitation and Characterization}\\
When using linearly polarized incident THz radiation, there are polarization selection rules for the excitation of the two magnon modes: the qFM mode is excited when a component of the THz magnetic field ($\mathbf{H}_{\mathrm{THz}}$) is perpendicular to the weak ferromagnetic moment ($\mathbf{F}$), and the qAFM mode is excited when a component of $\mathbf{H}_{\mathrm{THz}}$ is parallel to $\mathbf{F}$.\cite{Herrmann63JPCS} These selection rules are derived from solutions to the equations of motion for $\mathbf{S}_1$ and $\mathbf{S}_2$. To extend this analysis to the coupled modes, we numerically investigated the spin dynamics for the LM and UM. Notably, for $\theta = 90^\circ$, we found that the dynamics of the qFM mode when $\mathbf{H}_{\mathrm{DC}} = 0$ and the dynamics of the LM when $\mathbf{H}_{\mathrm{DC}} \neq 0$ are almost identical (see Supplementary Figs. 6a and 6h for a similar comparison). Given that the excitation of the former is forbidden by the selection rule, we also expect the latter to be forbidden.

In general, the transmitted THz electric fields are elliptically polarized\cite{YamaguchietAl13PRL} owing to the spin dynamics and the birefriengence of YFeO$_3$.\cite{JinetAl13PRB} In a tilted magnetic field, whether a magnon mode emits predominantly $\mathbf{E}_{\mathrm{THz}}^{a}$ or $\mathbf{E}_{\mathrm{THz}}^{b\textrm{-}c}$ polarized light is further complicated by the coupled spin dynamics and the angled cut of the crystal. Therefore, for a given $\theta$, whether $\mathbf{E}_{\mathrm{THz}}^{a}$ or $\mathbf{E}_{\mathrm{THz}}^{b\textrm{-}c}$ was used to characterize a magnon mode's frequency as a function of magnetic field depended on which polarization gave a larger signal. 

Transmitted THz electric fields at nonzero $\mathbf{H}_{\mathrm{DC}}$ are obtained by taking the difference between: (i) the signal measured at nonzero magnetic field and (ii) a reference, where the signal and reference are defined in the previous section. This yields a waveform in the time-domain, the Fourier transform of which reveals the magnon frequency at nonzero $\mathbf{H}_{\mathrm{DC}}$. This method of data analysis was used to characterize all magnon modes discussed in the main text, with the exception of the lower mode (LM) for $\theta = 60^\circ$. To study this weak oscillation, the subtracted reference was taken with THz transmitting through the YFeO$_3$ crystal at zero magnetic field, as opposed to without THz transmitting through the crystal. This method can be more sensitive because contributions from the large amplitude $\mathbf{E}_{\mathrm{THz}}$ pulse at $t=0$ are subtracted out. However, because the difference was taken between two THz waveforms that both propagated through the YFeO$_3$ crystal, one at zero $\mathbf{H}_{\mathrm{DC}}$ and one at nonzero $\mathbf{H}_{\mathrm{DC}}$, the analyzed THz waveform contains oscillations from magnons measured at both zero and at nonzero $\mathbf{H}_{\mathrm{DC}}$. This limits the ability to characterize the LM at nonzero $\mathbf{H}_{\mathrm{DC}}$ because its frequency will eventually overlap with the upper mode's (UM's) zero-field frequency. 

Supplementary Fig.\,3 shows the complete set of THz magnetospectroscopy measurements for each YFeO$_3$ crystal taken at applied field strengths up to $30$~T. Each plot indicates the $\theta$ and the magnon mode being identified. The spectra corresponding to different $\mathbf{H}_{\mathrm{DC}}$ are vertically offset with increasing field strength. The open circles indicate the qFM magnon frequency, the black circles indicate the UM frequency, and the black triangles indicate the LM frequency. The resonance frequencies and their corresponding magnetic fields are plotted in Figure~3a of the main text. The spectra were zero-padded for smoothing; the frequency resolution of the measurements ($1/T$ where $T$ is the Fourier-transformed time-range) is indicated by error bars in Figure~3a of the main text.

{\bf Equations of Motion and Hamiltonian}\\
We start from a microscopic spin model quantitatively describing interactions between the two Fe$^{3+}$ spin sublattices, including symmetric and antisymmetric exchange interactions, single-ion anisotropies, and the Zeeman interaction.\cite{Herrmann63JPCS} In the Supplementary Text, we derive the equations of motion in terms of $\mathbf{F} = \mathbf{R_1}+\mathbf{R_2}$ and $\mathbf{G} = \mathbf{R_1}-\mathbf{R_2}$, where $\mathbf{R_i}$ are the two spin-sublattice unit vectors. The equations of motion for small displacements of $\mathbf{F}$ and $\mathbf{G}$, given by $\delta \mathbf{F}$ and $\delta \mathbf{G}$ are derived as: 
\begin{equation} \label{eq:EOM_F_G}
    \begin{bmatrix}
        \delta \dot{F}_x\\
        \delta \dot{F}_y\\
        \delta \dot{G}_x\\
        \delta \dot{G}_y\\
    \end{bmatrix} = 2\gamma \mathrm{sin}\beta_z
    \begin{bmatrix}
        0 & 2A_y & D_{yx} & 0\\
        -2A_x & 0 & 0 & -D_{xy}\\
        D_{xy} & 0 & 0 & 2B_y\\
        0 & -D_{yx} & -2B_x & 0
    \end{bmatrix}
    \begin{bmatrix}
        \delta F_x\\
        \delta F_y\\
        \delta G_x\\
        \delta G_y\\
    \end{bmatrix}
\end{equation}
which yield two magnonic eigenfrequencies:
\begin{equation}
\begin{split}
    \omega_{\pm}^2 & = \frac{(4\gamma \mathrm{sin}\beta_z)^2}{2}\bigg( A_xA_y + B_xB_y - \frac{1}{2}D_{xy}D_{yx} \\ 
    & \pm \sqrt{(A_xA_y + B_xB_y - \frac{1}{2}D_{xy}D_{yx})^2 - 4(A_xB_y-\frac{1}{4}D_{xy}^2)(A_yB_x-\frac{1}{4}D_{yx}^2)}\bigg)
\end{split}
\end{equation}
where $\gamma$ is the gyromagnetic ratio, $\beta_z$ is the angle between $\mathbf{R_i}$ and the $a$-$b$ plane, and analytical expressions for $A_x$, $A_y$, $B_x$, $B_y$, $D_{xy}$, and $D_{yx}$ in terms of magnetic parameters are provided in the derivation in the Supplementary Text. 

One can show that when $\mathbf{H}_{\mathrm{DC}}$ is parallel to the $c$-axis, $D_{xy}$ and $D_{yx}$ exactly vanish. In this geometry, the equations of motion for $\delta F_{x,y}$ and $\delta G_{x,y}$ oscillations decouple, becoming the well-known qFM and qAFM modes,\cite{Herrmann63JPCS} respectively. We calculate the generalized, decoupled qFM and qAFM modes in a tilted magnetic field by neglecting $D_{xy}$ and $D_{yx}$. To calculate the magnon frequencies plotted in Figure~3a of the main text, we input magnetic parameters from previous studies of magnons in YFeO$_3$.\cite{KoshizukaHayashi88JPSJ}

To fully understand the magnonic interactions, we derive the following quantized Hamiltonian from the microscopic spin model in the Supplementary Text:
\begin{equation} \label{eq:H_ab}
    \mathcal{H} = \hbar\omega_{0a}\Big(\hat{a}^\dagger\hat{a}+\frac{1}{2}\Big)+\hbar\omega_{0b}\Big(\hat{b}^\dagger\hat{b}+\frac{1}{2}\Big)+i\hbar g_1\Big(\hat{a}\hat{b}^\dagger - \hat{a}^\dagger\hat{b}\Big)+i\hbar g_2\Big(\hat{a}^\dagger \hat{b}^\dagger - \hat{a}\hat{b}\Big)
\end{equation}
where $[\hat{a},\hat{a}^\dagger] = [\hat{b},\hat{b}^\dagger] = 1$, and $\omega_{0a}$ and $\omega_{0b}$ are the decoupled qFM and qAFM magnon frequencies in a general, tilted magnetic field. These are expressed in terms of the parameters in Eq. \ref{eq:EOM_F_G} as:
\begin{equation}
    \omega_{0a} = 4\gamma \mathrm{sin}\beta_z \sqrt{A_xA_y}
\end{equation}
\begin{equation}
    \omega_{0b} = 4\gamma \mathrm{sin}\beta_z \sqrt{B_xB_y}
\end{equation}
and $g_1$ and $g_2$ are the co-rotating and counter-rotating coupling strengths, respectively, expressed as:
\begin{equation} \label{eq:g1}
    g_1 = \gamma \mathrm{sin}\beta_z \bigg[ D_{xy} \big(\frac{A_yB_x}{A_xB_y}\big)^{1/4} - D_{yx}\big(\frac{A_xB_y}{A_yB_x}\big)^{1/4}\bigg]
\end{equation}
\begin{equation} \label{eq:g2}
    g_2 = \gamma \mathrm{sin}\beta_z \bigg[ D_{xy} \big(\frac{A_yB_x}{A_xB_y}\big)^{1/4} + D_{yx}\big(\frac{A_xB_y}{A_yB_x}\big)^{1/4}\bigg]
\end{equation}
The coupled magnon eigenfrequencies can be derived in terms of $g_1$, $g_2$, $\omega_{0a}$, and $\omega_{0b}$ from the equations of motion for the Hamiltonian in Eq. \ref{eq:H_ab}, which are provided in the Supplementary Text. These magnon eigenfrequencies are given by:
\begin{equation} \label{eq:Omegas}
    \Omega_{\pm}^2 = \frac{1}{2}[2g_1^2-2g_2^2+\omega_{0a}^2+\omega_{0b}^2\pm\sqrt{4g_1^2(\omega_{0a}+\omega_{0b})^2+(\omega_{0a}^2-\omega_{0b}^2)^2-4g_2^2(\omega_{0a}-\omega_{0b})^2}]
\end{equation}
where $\Omega_+$ ($\Omega_-$) is the UM (LM) eigenfrequency, and can be calculated and found to agree exactly with the previously calculated values, thereby confirming our quantized Hamiltonian.

{\bf Symmetry of Equations of Motion}\\
When $\mathbf{H}_{\mathrm{DC}}$ is applied along the $c$-axis, Eq.~(\ref{eq:EOM_F_G}) becomes block diagonal, and the eigenmodes are given by the well-known qFM and qAFM modes. This block diagonal matrix commutes with:
\begin{equation}
    \Sigma = \begin{bmatrix}
        -1 & 0 & 0 & 0\\
        0 & -1 & 0 & 0\\
        0 & 0 & 1 & 0\\
        0 & 0 & 0 & 1
    \end{bmatrix}
\end{equation}
which represents a $\pi$ rotation about the $c$-axis followed by sublattice exchange.\cite{MacNeilletAl19PRL} Therefore, the qAFM mode is unchanged under the operation of $\Sigma$, whereas the qFM mode gains a $\pi$ phase shift. However, when $\mathbf{H}_{\mathrm{DC}}$ is tilted in the $b$-$c$ plane, the $\pi$ rotational symmetry of $\mathbf{S}_{1}$ and $\mathbf{S}_{2}$ is broken, corresponding to nonzero $\beta_y$ in Supplementary Fig. 4. Accordingly, $\Sigma$ no longer commutes with the Eq.~(\ref{eq:EOM_F_G}) due to nonzero $D_{xy}$ and $D_{yx}$.

As described in the previous section, the generalized qFM and qAFM modes in a tilted magnetic field are defined as independent precessions of $\delta F_{x,y}$ and of $\delta G_{x,y}$, respectively, calculated in the absence of $D_{xy}$ and $D_{yx}$ in Eq.~(\ref{eq:EOM_F_G}). Thus, they are also eigenstates of $\Sigma$ and their parities (phases) are identical to those for the decoupled qFM and qAFM modes.

{\bf Spin Dynamics}\\
We numerically solve the equations of motion for $\mathbf{F}$ and $\mathbf{G}$ specified in Eq.~(\ref{eq:EOM_F_G}) for several geometries, which we transform back to the dynamics for the spin sublattice unit vectors $\mathbf{R_1}$ and $\mathbf{R_2}$. The dynamics for $\mathbf{R_i}$ take a simpler form when transformed from Cartesian coordinates ($X_i, Y_i, Z_i$) to the local, right-handed coordinate system ($S_i, T_i, Y_i'$) wherein $\mathbf{R_i}$ has components ($1, 0, 0$) in equilibrium. The transformation is illustrated in Supplementary Fig.\,4 and is discussed in the Supplementary Text. Supplementary Fig.\,5 shows the spin dynamics for $\theta = 0^\circ$ in the qFM and qAFM modes at applied field strengths of $5$~T and $20$~T (\textbf{a}-\textbf{d}), and for $\theta = 20^\circ$ in the qFM mode, qAFM mode, LM, and UM at an applied field strength of $5$~T (\textbf{e}-\textbf{h}). Supplementary Fig.\,6 displays the spin dynamics for $\theta = 90^\circ$ at applied field strengths of $5$~T and $20$~T in the qFM mode, qAFM mode, LM, and UM. The spin dynamics in each mode for an applied field strength of $5$~T are qualitatively illustrated in plots \textbf{c}, \textbf{f}, \textbf{i}, and \textbf{l}, with the position of each spin on its trajectory indicated in plots \textbf{a}, \textbf{d}, \textbf{g}, and \textbf{j}, respectively. 

{\bf Hopfield-Bogoliubov Transformation}\\
We perform a Hopfield-Bogoliubov transformation to diagonalize our Hamiltonian, Eq.~(\ref{eq:H_ab}). We introduce coupled magnon annihilation operators $\hat{B}_L$ $(\hat{B}_U)$ describing the LM (UM), which are expressed in terms of the generalized qFM (qAFM) operators $\hat{a}$ ($\hat{b}$) by:
\begin{equation} \label{eq:Bog}
    \hat{B}_j = W_j \hat{a} + X_j \hat{b} + Y_j \hat{a}^\dagger + Z_j \hat{b}^\dagger 
\end{equation}
for $j = L,U$. The coefficients are solutions to an eigenvalue problem discussed in the Supplementary Text. The Hamiltonian can be rewritten as: 
\begin{equation}
    \mathcal{H} = \hbar\Omega_- \hat{B}_L^\dagger \hat{B}_L + \hbar\Omega_+ \hat{B}_U^\dagger \hat{B}_U
\end{equation}
and the ground state $|0\rangle$ of our coupled magnon system must satisfy:
\begin{equation} \label{eq: gs}
    \hat{B}_j |0\rangle = 0
\end{equation}

{\bf Time-Reversed Components in Lower and Upper Modes}\\
As shown in Supplementary Fig.\,6, the qFM and LM precessions are almost identical. However, the major axes of the qAFM precessions are canted to the $T_{1,2}$ axes, while those of the UM are along the $Y'_{1,2}$ axes. Due to the presence of both co-rotating and the counter-rotating interactions, the coupled magnon dynamics should be a superposition of not only the decoupled qFM and qAFM modes, but also their time-reversals. In the following, we try to understand qualitatively how the time-reversed dynamics are included in the UM.


As seen in Supplementary Fig.\,6, the $Y'_{1,2}$ oscillations (dashed curves) are similar between the qAFM and UM. Thus, we must consider how the small $T_{1,2}$ oscillations (solid curves) in the UM are obtained by superposing the qAFM and qFM modes. The UM’s small $T_{1,2}$ oscillations ($\pi/2$ phase-shifted from $Y'_{1,2}$ oscillations) are already included in the qAFM. They are seen as the small left-shifted $T_1$ and right-shifted $T_2$ oscillations in the qAFM. Then, by eliminating the overall large oscillations of $T_{1,2}$ (roughly 0 or $\pi$ phase-shifted oscillation from $Y'_{1,2}$), we get the UM dynamics.

If we eliminate the qAFM’s overall $T_{1,2}$ oscillation simply by superposing the dynamics of the qFM, the qFM’s $Y'_{1,2}$ oscillations (dashed curves) are also added. They are in-phase with each other. However, the UM’s $Y'_{1,2}$ oscillations are out-of-phase with each other. So, the simple superposition of the qAFM and qFM cannot reproduce the UM dynamics.

The solution is the superposition not only with the qFM but also with the time-reversed qFM. The superposition of the qFM and its time-reversal ($1\to2\to3\to4$ and $3\to2\to1\to4$) has only the $T_{1,2}$ oscillations ($Y'_{1,2}$ oscillations are eliminated). Then, by superposing both the qFM and its time-reversal, the qAFM is transformed to the UM.

We quantitatively check the weight of the time-reversed qFM in the UM by evaluating the coefficients from the Hopfield-Bogoliubov transformation Eq.~(\ref{eq:Bog}). Supplementary Fig.\,7a shows the weights of the qFM ($|W_U|^2$) and qAFM ($|X_U|^2$) modes, and Supplementary Fig.\,7b shows those of the time-reversed qFM ($|Y_U|^2$) and time-reversed qAFM ($|Z_U|^2$) modes, all in the UM as functions of the applied field strength for $\theta=90^{\circ}$. The qFM mode ($|W_U|^2$) and its time-reversal ($|Y_U|^2$) have the same weight, consistent with the above discussion.

Supplementary Figs.\,7c and 7d show the same weights in the LM. While the spin dynamics of the qFM and LM are quite similar as discussed above, we observe that the LM also contains large time-reversed components. The $Y'_{1,2}$ amplitudes in the LM are in fact slightly larger (about 4\;\%) than those in the qFM. The time-reversed qFM and qAFM are required for reproducing this difference.

{\bf Squeezing}\\
To demonstrate that $|0\rangle$ satisfying $\hat{B}_L |0\rangle = \hat{B}_U |0\rangle = 0$ is an intrinsically quantum vacuum squeezed state, we first introduce two orthogonal, generalized annihilation operators:
\begin{equation}
    \hat{c} = \alpha \hat{a} + \beta \hat{b}
\end{equation}
\begin{equation}
    \hat{d} = \beta^* \hat{a} - \alpha \hat{b}
\end{equation}
where $\alpha \in \mathbb{R}$, $\beta \in \mathbb{C}$ and they satisfy $\alpha^2 + |\beta|^2 = 1$. With respect to these generalized annihilation operators, we define the following quadratures:
\begin{equation}
    \hat{X}_{\hat{c},\phi} = (\hat{c}e^{i\phi}+\hat{c}^\dagger e^{-i\phi})/2
\end{equation}
\begin{equation}
    \hat{X}_{\hat{d},\phi} = (\hat{d}e^{i\phi}+\hat{d}^\dagger e^{-i\phi})/2
\end{equation}
The standard quantum limit for both of these operators is given by $1/4$. 

The variances of $\hat{X}_{\hat{c},\phi}$ and $\hat{X}_{\hat{d},\phi}$ can be easily evaluated in $|0\rangle$ by inverting the Hopfield-Bogoliubov transformation and rewriting the quadratures in terms of $\hat{B}_j$. Expressions for these variances are provided in the Supplementary Text. Using these expressions, we minimized the quadrature variances by numerically searching for the optimal $\alpha$, $\beta$, and $\phi$. Supplementary Fig.\,8 shows the fluctuation suppression for $\theta = 20^\circ$, $40^\circ$, $60^\circ$, and $90^\circ$. Note that we did not observe squeezing for the case when $\theta = 0^\circ$. 


{\bf Squeezing and Phase Transition}\\
To evaluate the contribution of $g_2$ to the squeezing,
we numerically calculated the minimum quadrature variance
$\langle 0|(\Delta \hat{X}_{\hat{c},\phi})^2 |0\rangle$
while artificially changing $|g_2|$
and keeping the other parameters
as obtained at $\theta = 90^{\circ}$ and $30$~T.
In Supplementary Fig.\,9a, the minimum $\langle 0|(\Delta \hat{X}_{\hat{c},\phi})^2 |0\rangle$ is plotted
as a function of $|g_2|$.
The minimum quadrature variance is increased to $0.25$ (0\;dB) at $|g_2| = 0$.
By increasing $|g_2|$, one can find that the minimum 
quadrature variance drops to zero at $|g_2| = 2\pi\times0.763\;\mathrm{THz}$.

This condition corresponds to the superradiant phase transition when we transform our anisotropic Hopfield Hamiltonian,
Eq.~(\ref{eq:H_ab}), into the anisotropic Dicke Hamiltonian, given by:
\begin{equation}
    \mathcal{H} \to \hbar\omega_{0a}\left(\hat{a}^\dagger\hat{a}+\frac{1}{2}\right)+\hbar\omega_{0b}\left(\hat{S}_z + \frac{N}{2}\right)+\frac{i\hbar g_1}{\sqrt{N}}(\hat{a}\hat{S}_+ - \hat{a}^\dagger\hat{S}_-)+\frac{i\hbar g_2}{\sqrt{N}}(\hat{a}^\dagger \hat{S}_+ - \hat{a}\hat{S}_-)
\end{equation}
Here, $\hat{S}_{x,y,z}$ are the spin-$\frac{N}{2}$ operator,
and $\hat{S}_{\pm} \equiv \hat{S}_x \pm i \hat{S}_y$
are the raising and lowering operators.
The phase transition is obtained
in this Hamiltonian when the LM's eigenfrequency becomes zero,
indicating an instability of the normal phase.
This condition is derived from Eq.~\ref{eq:Omegas} as:
\begin{equation} \label{eq:anisotropic_SRPT}
    1 + \frac{(g_1{}^2-g_2{}^2)^2}{\omega_{0a}{}^2\omega_{0b}{}^2} - \frac{2(g_1{}^2+g_2{}^2)}{\omega_{0a}\omega_{0b}} = 0
\end{equation}
In the isotropic case $g_1 = g_2 = g$, this is reduced to the well-known condition $4g^2 = \omega_{0a}\omega_{0b}$ of the superradiant phase transition in the isotropic Dicke Hamiltonian.

The drop condition $|g_2| = 2\pi\times0.763\;\mathrm{THz}$
of the minimum quadrature variance in Supplementary Fig.\,9a
satisfies Eq.~\ref{eq:anisotropic_SRPT}.
In this way, the minimum quadrature variance becomes
zero at the superradiant phase transition.

{\bf Comparison of VBSS and VRSS}\\
We define the VBSS and VRSS for the UM as:
\begin{equation}
    \mathrm{VBSS} = \Omega_+(g_1 \neq 0, g_2 = 0) - \Omega_+(g_1 \neq 0, g_2 \neq 0)
\end{equation}
\begin{equation}
    \mathrm{VRSS} = \Omega_+(g_1 \neq 0, g_2 = 0) - \mathrm{max}(\omega_{0a},\omega_{0b})
\end{equation}
Here, we assume that $\mathrm{max}(\omega_{0a},\omega_{0b}) = \omega_{0b}$, but similar results can be derived for $\mathrm{max}(\omega_{0a},\omega_{0b}) = \omega_{0a}$.

The condition for $\mathrm{VBSS} > \mathrm{VRSS}$ is derived from Eq.~\ref{eq:Omegas} as:
\begin{equation} \label{VBSS_VRSS}
    2\omega_{0a}\omega_{0b}(g_1^2 + g_2^2)
    < 
    (g_2^2 - g_1^2)\big[ (g_2^2-g_1^2) + 2\omega_{0b}^2 \big]
\end{equation}
We can immediately identify that this cannot be satisfied in the isotropic case where $g_1^2 = g_2^2$. The condition for the normal phase, from Eq. \ref{eq:anisotropic_SRPT}, is given by:
\begin{equation}
    g_1^2 - g_2^2 < \omega_{0a}\omega_{0b} - \sqrt{4g_2^2\omega_{0a}\omega_{0b}}
\end{equation}
Under the assumption that $|g_1| > |g_2|$, one can only satisfy Eq. \ref{VBSS_VRSS} if:
\begin{equation}
    g_1^2 - g_2^2 > \omega_{0b}(\omega_{0a}+\omega_{0b})+\sqrt{4g_2^2\omega_{0a}\omega_{0b}+\omega_{0b}^2(\omega_{0a}+\omega_{0b})^2}
\end{equation}
Thus, for $|g_1| > |g_2|$, one cannot achieve $VBSS > VRSS$ for the UM in the normal phase. However, for $|g_2| > |g_1|$, the condition for $\mathrm{VBSS} > \mathrm{VRSS}$ can be derived as:
\begin{equation}
g^{ 2 }_{2}  > g^{2}_{1} - \omega_{0b} ( \omega_{0b} - \omega_{0a} ) + \sqrt{ \omega^{2}_{0b} ( \omega_{0b} - \omega_{0a} )^{ 2 } + 4 \omega_{0a} \omega_{0b} g^{2}_{1} }  \quad ( > g^{2}_{1} )
\end{equation}
which can be satisfied in the normal phase. 




{\bf Discontinuity for $\theta = 90^\circ$}\\
To calculate normalized coupling strengths presented in Figure~4c of the main text, we require the magnetic field $H_{\mathrm{cross}}$ at which the generalized qFM and qAFM mode frequencies cross. Supplementary Fig.\,10a plots calculated values of $H_{\mathrm{cross}}$ for all $0^\circ \leq \theta \leq 90^\circ$. 

We observe that at an applied field strength of $1,284$~T for $\theta = 90^\circ$, a magnetically-driven phase transition occurs, where $\mathbf{S_1}$ and $\mathbf{S_2}$ become perfectly aligned along the $b$-axis. We also find that the generalized qFM and qAFM magnon frequencies, as well as the coupling strengths $g_1$ and $g_2$, are unstable at this point and change discontinuously, leading to a discontinuity in the normalized coupling strengths, illustrated in Supplementary Fig.\,10b.


\newpage


\item [{\bf \,\,\,Acknowledgements:}] We thank Kaden Hazzard and Han Pu for useful discussions. We thank Kevin Tian for assistance with measurements and Tanyia Johnson for the illustration of the pulsed magnet system. J.K.\ acknowledges support from the Army Research Office (Grant No.\ W911NF-17-1-0259). S.C.\ acknowledges support from the National Natural Science Foundation of China (NSFC, No.\ 11774217).

\item[{\bf \,\,\,Author contributions:}] T.M.\ performed all measurements and analyzed all experimental data under the supervision and guidance of G.T.N.\ and J.K. X.L.\ and N.M.P.\ assisted T.M.\ in measurements and data analysis. K.H.\ calculated resonance frequencies, coupling strengths, and the degree of squeezing. M.B.\ proposed the control of coupling strengths by changing the magnetic field orientation and supervised T.M.\ and K.H.\ in the theoretical analysis. X.M.\ grew, cut, and characterized the crystals used in the experiments under the guidance of W.R., G.M., and S.C. I.K.\ and J.T.\ conceived the single-shot detection technique. H.N.\ developed the 30-T pulsed magnet. Z.J.\ and D.T.\ characterized the THz response of the samples at low magnetic fields. T.M., K.H., M.B., and J.K. wrote the manuscript. All authors discussed the results and commented on the manuscript.
\item[{\bf \,\,\,Correspondence:}] Correspondence and requests for materials should be addressed to Shixun Cao (email: sxcao@shu.edu.cn), Motoaki Bamba (email: bamba.motoaki.8a@kyoto-u.ac.jp), and Junichiro Kono (email: kono@rice.edu).
\item[{\bf \,\,\,Supplementary Information:}] \textbf{Supplementary Information} is available for this paper.



\end{scilastnote}

\end{document}